 \documentclass[final,3p,times,twocolumn]{elsarticle}

 \usepackage{graphicx}

\usepackage{amssymb}
\usepackage{color}

\usepackage{siunitx}
\newcommand{\nuc}[2]{\hbox{$^{#1}$#2}}
\usepackage{gensymb}
\DeclareSIUnit[number-unit-product = {}]{\inchQ}{\textquotedbl}
\DeclareSIUnit[number-unit-product = {\thinspace}]{\inch}{in}

\usepackage[colorlinks]{hyperref}

\usepackage{dcolumn}
\usepackage{bm}
\usepackage{float}
\usepackage{multirow}
\usepackage{longtable}
\usepackage{dcolumn}
\newcolumntype{d}{D{.}{.}{-1}}
\usepackage{setspace}
\usepackage{threeparttable}
\usepackage{tabularx}
\usepackage[symbol*]{footmisc}
\DefineFNsymbolsTM{myfnsymbols}{
  \textasteriskcentered *
  \textdagger    \dagger
  \textdaggerdbl \ddagger
  \textsection   \mathsection
  \textbardbl    \|%
  \textparagraph \mathparagraph
}%
\setfnsymbol{myfnsymbols}

\usepackage{gensymb}

\biboptions{sort,compress}
\journal{Physics Letters B}

\begin{document}

\begin{frontmatter}



\title{Hexadecapole strength in the rare isotopes \nuc{74,76}{Kr}}


\author[fsu]{M.~Spieker\corref{cor1}}
\ead{mspieker@fsu.edu}
\cortext[cor1]{Corresponding author}
\address[fsu]{Department of Physics, Florida State University, Tallahassee, FL 32306, USA}

\author[frib]{S.E. Agbemava}
\address[frib]{Facility for Rare Isotope Beams, Michigan State University, East Lansing, Michigan 48824, USA}

\author[frib,msu]{D. Bazin}
\address[msu]{Department of Physics and Astronomy, Michigan State University, East Lansing, Michigan 48824, USA}

\author[frib]{S. Biswas}

\author[fsu]{P.D. Cottle}

\author[frib,msu]{P.J. Farris}

\author[frib,msu]{A. Gade}

\author[frib]{T. Ginter}

\author[frib]{S. Giraud}

\author[fsu]{K.W. Kemper}

\author[frib]{J. Li}

\author[frib,msu]{W. Nazarewicz}

\author[frib]{S. Noji}

\author[frib]{J. Pereira}

\author[ursinus]{L.A. Riley}
\address[ursinus]{Department of Physics and Astronomy, Ursinus College, Collegeville, PA 19426, USA}

\author[frib]{M. Smith}

\author[frib]{D. Weisshaar}

\author[frib,msu]{R.G.T. Zegers}

\begin{abstract}
In the Ge-Sr mass region, isotopes with neutron number $N \leq 40$ are known to feature rapid shape changes with both nucleon number and angular momentum. To gain new insights into their  
structure, inelastic proton scattering experiments in inverse kinematics were performed on the rare isotopes \nuc{74,76}{Kr}. This work focuses on observables related to the $J^{\pi} = 4^+_1$ states of the Kr isotopes and, in particular, on the hexadecapole degree of freedom. By performing coupled-channels calculations,  hexadecapole deformation parameters $\beta_4$ were determined for the $J^{\pi} = 4^+_1$ states of \nuc{74,76}{Kr} from inelastic proton scattering cross sections. Two possible coupled-channels solutions were found. A comparison to predictions from nuclear energy density functional theory, employing both non-relativistic and relativistic functionals, clearly favors the large, positive $\beta_4$ solutions. These  $\beta_4$ values are unambiguously linked to the well deformed prolate configuration. Given the  $\beta_2 - \beta_4$ trend, established in this work, it  appears that $\beta_4$ values could provide a sensitive measure
of the nuclear shell structure. 
\end{abstract}


\end{frontmatter}



The neutron-deficient, even-even Ge, Se, Kr, and Sr isotopes exhibit rapid shape changes with both nucleon number and spin of the system \cite{ZrBook}. Generally, the ground-state shapes appear to change from prolate to oblate towards the end of the deformed region \cite{Naz96a,Jol03a,Yan21a}.
Current experimental data suggest that the prolate-oblate ground-state shape transition for even-$A$ nuclei in the Ge-Kr region occurs around neutron number $N = 36$ \cite{Lec80b, Gad05a, Lju08a, Iwa14a, Hen18a, Wim20a}. The exact location of the shape transition is still under debate and its details challenge state-of-the-art theoretical models as triaxial degrees of freedom are expected to contribute \cite{Aya16a, Hen19a}. Additional complexity gets added as nuclei in this mass region display complex shape coexistence of oblate, prolate, spherical and triaxial configurations at low excitation energy \cite{Hey11a, Gar21a}. Many  models (see Refs. \cite{Ben06a, Gir09a, Hin10a, Sat11a, Rod14a}) predict predominantly oblate-deformed ground states for nuclei around $N = 36$, with the yrast structure changing from oblate to prolate with angular momentum. This has been 
discussed in Ref.\,\cite{Iwa14a}, which showed that the models that incorporate mixing between oblate and prolate configurations  were able to describe both the experimentally measured $B(E2; 2^+_1 \rightarrow 0^+_1)$ and $B(E2; 4^+_1 \rightarrow 2^+_1)$ strengths in \nuc{72}{Kr}. Interestingly, a good description of the $B(E2; 2^+_1 \rightarrow 0^+_1)$ strength in \nuc{72}{Kr} has been offered by several models while predicting different ground-state structures \cite{Iwa14a}. The $4^+_1$ state and associated observables as, {\it e.g.}, $\gamma$-decay probabilities or excitation cross sections could, therefore, be extremely effective discriminators between competing model descriptions.

To further investigate the sensitivity of observables associated with the $4^+_1$ state to the structure of nuclei in the Ge-Sr mass region, we report on  $B(E4; 4^+_1 \rightarrow 0^+_1)$ strengths and associated  $\beta_4$ hexadecapole deformation parameters  in \nuc{74,76}{Kr}. These quantities were derived from inelastic proton scattering cross sections measured in inverse kinematics at the Coupled Cyclotron Facility of the National Superconducting Cyclotron Laboratory (NSCL) at Michigan State University\,\cite{NSCL} with the NSCL/Ursinus Liquid Hydrogen (LH$_2$) Target, the GRETINA $\gamma$-ray tracking array \cite{Pas13a,Wei17a}, and S800 magnetic spectrograph \cite{Baz03}. Data from low-energy Coulomb excitation (CoulEx) experiments had previously provided strong evidence for  prolate ground states but also indicated significant mixing between prolate and oblate configurations \cite{Cle07a}. Using two-state mixing arguments,  a large quadrupole deformation has been attributed to  the prolate configuration \cite{Cle07a}. As will be discussed in this 
Letter,  large intrinsic quadrupole deformations of the prolate configurations are  consistent with the    hexadecapole deformation parameters for \nuc{74,76}{Kr}, thus supporting the sensitivity of  $\beta_4$ to the quadrupole shape of the nucleus.

\begin{figure}[t]
\centering
\includegraphics[width=1\linewidth]{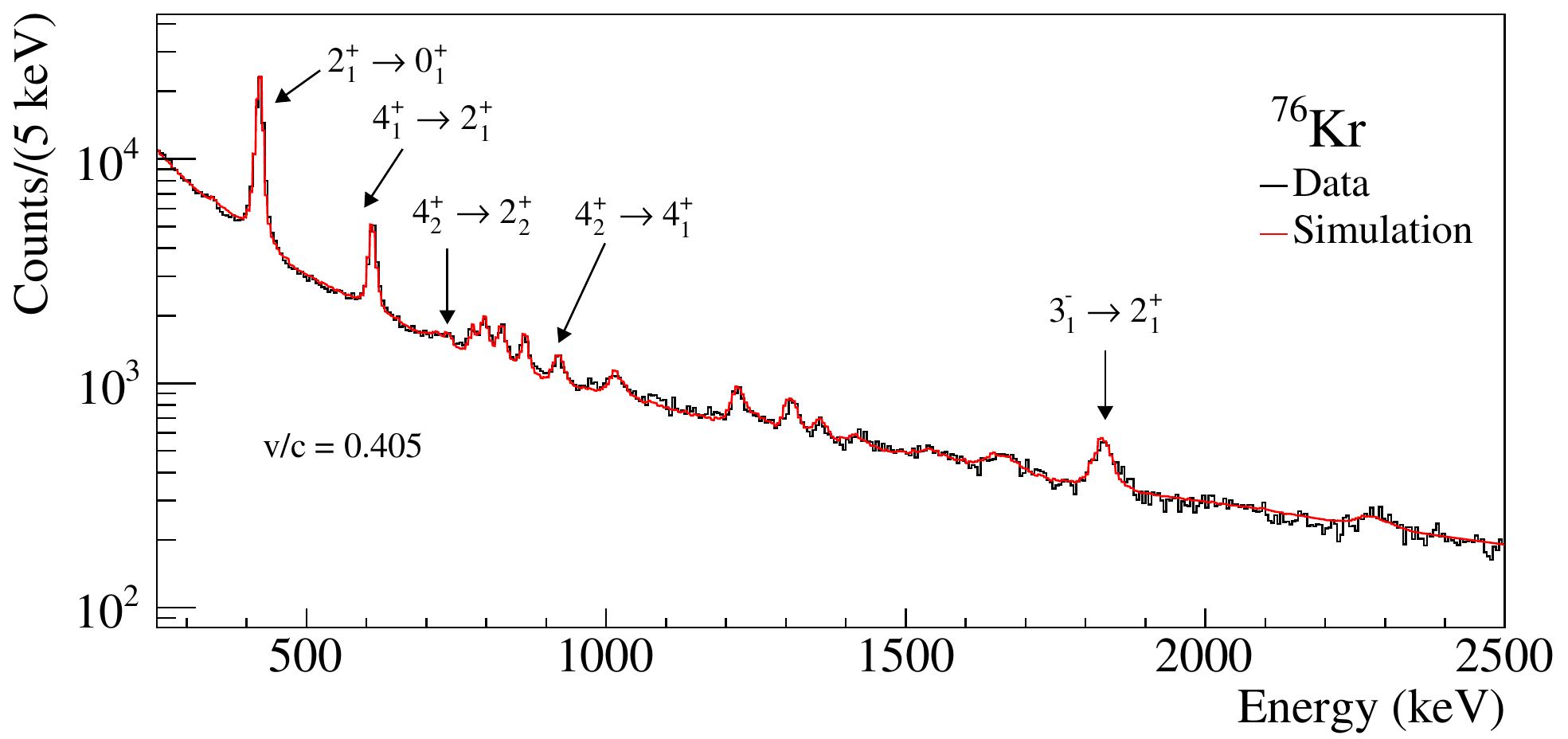}
\includegraphics[width=1\linewidth]{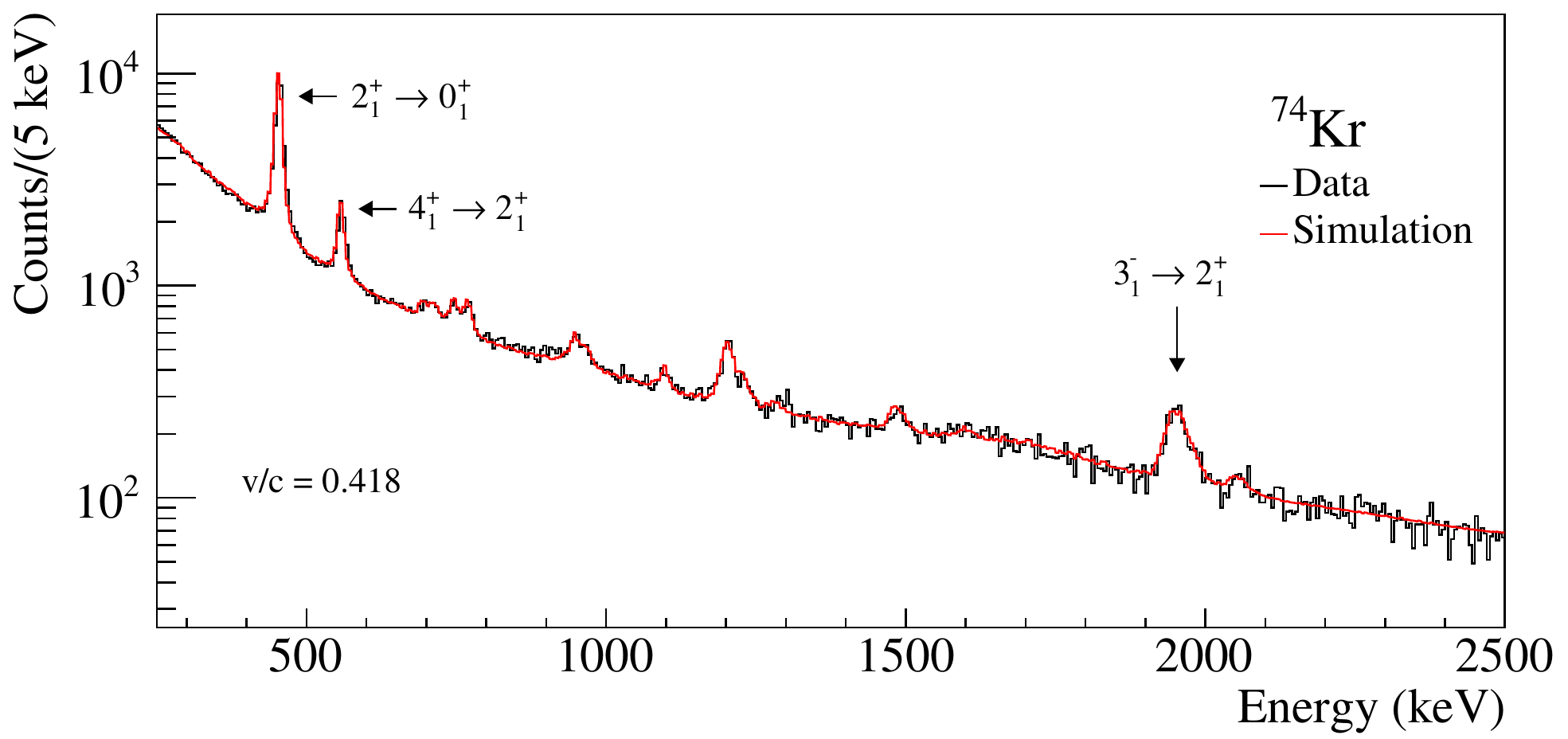}
\caption{\label{fig:spectra}{Doppler-corrected, in-beam $\gamma$-ray spectra for \nuc{76}{Kr} (top) and \nuc{74}{Kr} (bottom). Data are shown in black. \textsc{geant4} simulations performed with \textsc{ucgretina} \cite{Ril21a} are presented in red. A prompt background consisting of two exponential functions was included in the simulation. Besides transitions from the $J^{\pi} = 2^+_1$ and $3^-_1$ states, observed transitions of the populated $4^+$ states are highlighted.}} 
 \end{figure}

\begin{figure}[t]
\centering
\includegraphics[width=1\linewidth]{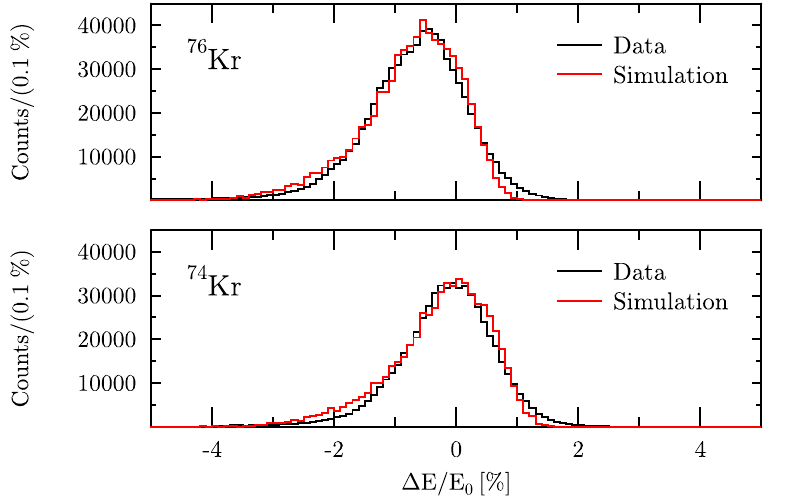}
\caption{\label{fig:dta}{Measured kinetic-energy distributions of the outgoing \nuc{74,76}{Kr} beams (black) compared to results of detailed \textsc{geant4} simulations (red) performed with \textsc{ucgretina} \cite{Ril21a} to determine the LH$_2$ target thickness. Both simulations provided consistent results for the target thickness.}} 
\end{figure}

For the experiments, the secondary \nuc{76}{Kr} (79\,$\%$ purity; $\sim 4740$ pps) and \nuc{74}{Kr} (51\,$\%$ purity; $\sim 3060$ pps) beams were produced from a 150\,MeV/u $^{78}$Kr primary beam in projectile fragmentation on a 308-mg/cm$^2$ thick $^{9}$Be target. To select \nuc{74,76}{Kr} in flight, two separate A1900 magnetic settings and a 240-mg/cm$^2$ Al degrader were used. Both secondary beams were unambiguously distinguished from other components in the cocktail beam via the time-of-flight difference measured between two plastic scintillators located at the exit of the A1900 and the object position of the S800 analysis beam line. At the chosen proton center-of-mass energies of about 100 MeV, both proton and neutron matrix elements are probed  \cite{Cot01a}. The NSCL/Ursinus Liquid Hydrogen (LH$_2$) Target with the 8.5-mm thick cell was installed at the target position of the S800 spectrograph. The S800 magnetic spectrograph and its focal-plane detection system were used to identify the projectile-like reaction residues event-by-event from their energy loss and time of flight \cite{Baz03}. $\gamma$ rays, which were emitted by the reaction residues in flight, were detected with the GRETINA $\gamma$-ray tracking array \cite{Pas13a,Wei17a}. Eight GRETINA modules, containing four, 36-fold segmented HPGe detectors each, were mounted in the north half of the mounting shell to accommodate the LH$_2$ target. In this configuration, two modules are centered at 58$^{\circ}$, four at 90$^{\circ}$, and two at 122$^{\circ}$ with respect to the beam axis. At beam velocities of $v/c \approx 0.4$, event-by-event Doppler reconstruction of the residues' $\gamma$-ray energies is necessary. This reconstruction was performed based on the angle of the $\gamma$-ray emission determined from the main-interaction point and including trajectory reconstruction of the residues through the S800 spectrograph \cite{Wei17a}. Doppler-corrected in-flight $\gamma$-ray spectra are presented in Fig.\,\ref{fig:spectra}. The $\gamma$-ray yields were obtained by fitting $\gamma$-ray spectra, simulated with \textsc{ucgretina} \cite{Ril21a}, to the experimentally observed ones. For these fits, the \textsc{root} \cite{root} \textsc{minuit2} minimizer with the default minimization algorithm \textsc{migrad} was used \cite{minuit2}; as done in previous studies, see, {\it e.g.}, Refs.\,\cite{Ril19a, Hil21a, Spi22a}. Known decay branching for excited states of \nuc{74,76}{Kr} \cite{ENSDF, Gia05a, Dun13a} was explicitly taken into account by using the \textsc{geant4} photo-evaporation database format \cite{geant4} implemented in \textsc{ucgretina}. This procedure also allowed for the correction of the $\gamma$-ray yields for observed feeders (see also Ref.\,\cite{Spi22a}). Inelastic scattering cross sections were calculated from the experimental $\gamma$-ray yields by normalizing these to the number of incoming beam particles and the number of target nuclei. Pressure differences across the Kapton entrance and exit windows of the LH$_2$ cell cause them to bulge outwards. As first described in Ref.\,\cite{Ril19a}, the LH$_2$ target thickness was determined via a comparison of the measured kinetic-energy distribution of the reacted outgoing beam to a detailed \textsc{geant4} simulation performed with \textsc{ucgretina} \cite{Ril21a}. The simulation also uses the independently measured kinetic-energy distribution of the incoming beam through the empty target cell as input. The comparison is shown in Fig.\,\ref{fig:dta}. Excellent agreement was obtained and an areal target density of 69(3) mg/cm$^2$ was determined.

Reaction calculations were performed with the coupled-channels program \textsc{chuck3} \cite{chuck} using the optical-model parameters of \cite{Kon03a}. For one-step processes, deformation parameters $\beta_{\lambda}$ can be calculated by scaling the theoretical to the experimentally determined inelastic scattering cross sections. Excellent agreement with the adopted values \cite{Pri14a, Pri16a} was obtained for the $\beta_2$ quadrupole deformation parameters of the $2^+_1$ states, $\beta_2 = 0.40 \pm 0.02 \ \mathrm{(stat.)} \pm 0.03 \ \mathrm{(sys.)}$ for \nuc{76}{Kr} and $\beta_2 = 0.35 \pm 0.06 \ \mathrm{(stat.)} \pm 0.02 \ \mathrm{(sys.)}$ for \nuc{74}{Kr}, benchmarking both the reaction calculations and feeding correction (see also Ref.\,\cite{Spi22a}). For possible multi-step processes, we followed the approach of Ref. \cite{Sak79a} for the stable Kr isotopes. In particular, coupled-channels calculations were performed to determine the hexadecapole deformation parameter, $\beta_4$, from the inelastic scattering cross sections measured for the $4^+_1$ states of \nuc{74,76}{Kr}. The inelastic scattering cross sections to the $4^+_1$ state are 5.1(5)\,mb for \nuc{76}{Kr} and 6.1(11)\,mb for \nuc{74}{Kr}, respectively. Additional feeding contributions coming from unobserved feeders cannot be ruled out entirely. For the two-step contribution via the intermediate $2^+_1$ state, the adopted $\beta_2$ values were chosen \cite{Pri14a, Pri16a}. In principle, this leaves $\beta_4$ as the only free parameter to match  the experimental cross section. Using this approach, values of $\beta_4 = 0.201 \pm 0.009 \ \mathrm{(stat.)} \pm 0.016 \ \mathrm{(sys.)}$ for \nuc{76}{Kr} and $\beta_4 = 0.23 \pm 0.02 \ \mathrm{(stat.)} \pm 0.02 \ \mathrm{(sys.)}$ for \nuc{74}{Kr} were determined. Systematic uncertainties include the uncertainty of the adopted $\beta_2$ value \cite{Pri14a, Pri16a} and uncertainties in the reaction kinematics due to the target thickness. Systematic uncertainties due to unobserved feeders cannot be estimated reliably. It is, however, important to state that if there existed a single feeder or even several feeders leading to a further feeding-subtracted $\beta_4$ value of around $+0.1$ as in \nuc{78}{Kr} \cite{Sak79a}, then we should have been able to observe the corresponding $\gamma$-ray transitions up to an energy of at least 2\,MeV. We also note that there exist alternative coupled-channels solutions corresponding to  negative $\beta_4$ values. These values are $\beta_4 = -0.127 \pm 0.009 \ \mathrm{(stat.)} \pm 0.022 \ \mathrm{(sys.)}$ for \nuc{76}{Kr} and $\beta_4 = - 0.17 \pm 0.02 \ \mathrm{(stat.)} \pm 0.02 \ \mathrm{(sys.)}$. Positive and negative $\beta_4$ solutions from inelastic scattering cross sections were also reported in the rare-earth region \cite{Ron77a}; they correspond to constructive and destructive interference between one-step and two-step processes.
In this context, we want to  stress that two-step excitation alone would only provide cross sections of $\sim 0.8$\,mb for \nuc{76}{Kr} and $\sim 0.5$\,mb for \nuc{74}{Kr}, respectively. A significant direct (one-step) contribution is therefore needed to explain the experimental data. As we will discuss later, the positive solutions for $\beta_4$ are preferred. We note, however,  that the measured inelastic scattering cross sections have no sensitivity to the sign of  $\beta_4$. Therefore, both values, which were determined through the coupled-channels analysis, are reported in this Letter.

\begin{figure}[t]
\centering
\includegraphics[width=1\linewidth]{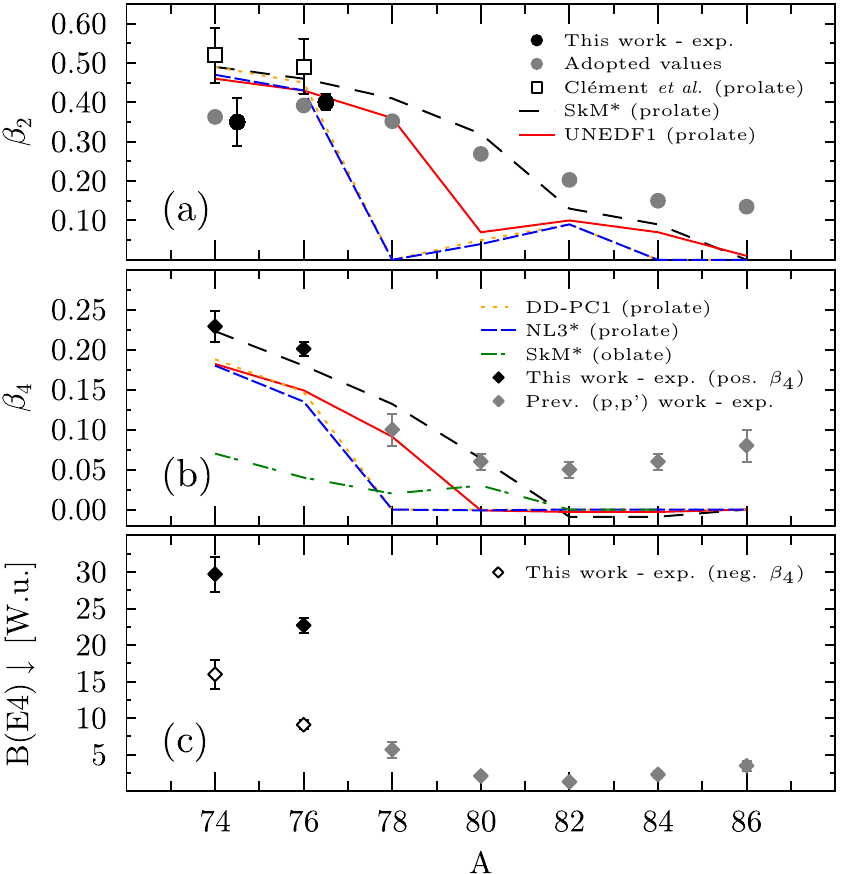}
\caption{\label{fig:sys}{(a) Experimental $\beta_2$ values [new data (black circles), adopted values \cite{Pri14a, Pri16a} (gray circles), and values determined by Cl{\'e}ment {\it et al.} for the unperturbed prolate configuration \cite{Cle07a} (white squares)], (b) $\beta_4$ values [new data (black diamonds) and previous $(p,p')$ data \cite{Sak79a} (gray diamonds)], and (c) experimental $\mathrm{B}(\mathrm{E4};4^+_1 \rightarrow 0^+_1)$ strengths in neutron-deficient Kr isotopes. Alternative values, obtained for the negative $\beta_4$ values discussed in the text, are presented with white diamonds in panel (c). Theoretical DFT predictions for the corresponding observables are shown in panels (a) and (b). These were obtained using the energy density functionals SkM* (black longer dashed line) and UNEDF1 (red solid line), as well as two covariant energy density functionals DD-PC1 (orange dotted line) and NL3* (blue short dashed line). For $\beta_2$, predictions are shown for the prolate minimum only. In addition, predictions for the oblate minimum made with SkM* are shown (green dotted-dashed line) in (b).}} 
 \end{figure}

The $\beta_2$ deformation parameters determined in our experiments \cite{Spi22a}, the adopted values \cite{Pri14a, Pri16a}, as well as the calculated $\beta_2$ values for the unperturbed prolate configuration are shown in Fig.\,\ref{fig:sys}\,(a). The latter were determined by Cl{\'e}ment {\it et al.} through a two-state mixing calculation as described in Ref.\,\cite{Cle07a}. The positive $\beta_4$  deformation parameters for  \nuc{74,76}{Kr} are shown in Fig.\,\ref{fig:sys}\,(b), which also includes the  data of Ref.\,\cite{Sak79a} for isotopes with $A \geq 78$. A significant increase of $\beta_4$ is observed in \nuc{74,76}{Kr}. Fig.\,\ref{fig:sys}\,(c) shows the evolution of the $B(E4;4^+_1 \rightarrow 0^+_1)$ strength along the Kr isotopic chain. Consistent with Ref.\,\cite{Sak79a}, the $B(E4)$ strengths for \nuc{74,76}{Kr} were calculated from the $\beta_4$ parameters by assuming a uniform mass distribution of radius $R= 1.2 \ \mathrm{fm} \times A^{1/3}$ \cite{Ber69a}. Using the positive $\beta_4$ values, significant $B(E4;4^+_1 \rightarrow 0^+_1)$ strengths of $22.7 \pm 1.0 \ \mathrm{(stat.)} \pm 1.8 \ \mathrm{(sys.)}$ W.u. for \nuc{76}{Kr} and $29 \pm 2 \ \mathrm{(stat.)} \pm 2 \ \mathrm{(sys.)}$ W.u. for \nuc{74}{Kr} are determined. The $B(E4)$ strengths, determined from the negative $\beta_4$ solutions, are also shown in  Fig.\,\ref{fig:sys}\,(b). The $B(E4; 4^+_1 \rightarrow 0^+_1)$ strengths (and $\beta_4$ values) of the corresponding Se and Ge isotones are significantly smaller \cite{Ogi86a, Sch87a}. It is worth noting that strengths larger than 10 W.u. were also observed in the $A=100$ region \cite{Pig92a} as well as in the Nd isotopes \cite{Pig93a} in inelastic light-ion scattering experiments. Besides the $J^{\pi} = 4^+_1$ state of \nuc{76}{Kr}, the 1957-keV, $J^{\pi} = 4^+_2$ state was also populated in our $(p,p')$ experiment (see Fig.\,\ref{fig:spectra}). As this state does not belong to the ground-state band of \nuc{76}{Kr}, single-step excitation was assumed and $\beta_4 = 0.151 \pm 0.011 \ \mathrm{(stat.)} \pm 0.012 \ \mathrm{(sys.)}$ derived. This $\beta_4$ value corresponds to a $B(E4)$ strength of $12.9 \pm 0.9 \ \mathrm{(stat.)} \pm 2.0 \ \mathrm{(sys.)}$ W.u., which brings the total $B(E4)$ strength up to $36 \pm 2 \ \mathrm{(stat.)} \pm 4 \ \mathrm{(sys.)}$ W.u. in \nuc{76}{Kr}. No higher-lying, excited $4^+$ states were observed for \nuc{74}{Kr}. In the following, we concentrate on the $J^{\pi} = 4^+_1$ states of the Kr isotopes and associated observables.

To investigate the origin of the large $\beta_4$ values in \nuc{74,76}{Kr}, nuclear density functional theory (DFT) calculations were performed  using the Skyrme SkM*~\cite{BARTEL198279} and UNEDF1~\cite{Kortelainen2012} energy density functionals, as well as the covariant NL3*~\cite{LALAZISSIS2009} and DD-PC1~\cite{Niksic2008} energy density functionals. As in Ref.\,\cite{Stoitsov2003}, the mixed-type density-dependent delta interaction ~\cite{Dobaczewski2002} with the Lipkin-Nogami approximate particle-number projection was used in the pairing channel for the Skyrme functionals (SEDFs). The separable finite-range pairing~\cite{TIAN200944} with the strength defined as in Ref.\,\cite{Agbemava2014} was adopted for the covariant functionals (CDFTs). For the SEDFs, calculations were performed with the parallel DFT axial solver HFBTHO\,\cite{PEREZ2017363}. For the CDFTs, the calculations were carried out employing an axial solver of
Ref.\,\cite{Agbemava2014}. 

The charge multipole  moments $Q_{\lambda 0}$ ($\lambda=2, 4$) and dimensionless deformation
parameters $\beta_{\lambda}$ are defined as:
\begin{equation}\label{beta_def}
Q_{\lambda 0}=\sqrt{\frac{16\pi}{2\lambda+1}}\langle r^\lambda Y_{\lambda 0}\rangle=
 \frac{3}{\sqrt{\left( 2 \lambda + 1 \right) \pi}} Z R_0^{\lambda} \beta_{\lambda},
\end{equation}
where $R_0 = 1.2A^{{1}/{3}}$. For strongly deformed nuclei, deformation parameters $\beta_\lambda$
defined through the linear relation (\ref{beta_def}) can strongly deviate from commonly used deformation parameters entering the multipole expansion of the nuclear surface \cite{Naz96a, Lea88a}. Consequently, one has to exercise caution when comparing to experimental data.

The  predicted trend  of $\beta_4$ with mass number $A$ is shown in Fig.\,\ref{fig:sys}(b). Keeping the linear approximation of Eq.\,(\ref{beta_def}) in mind, all our models predict a significant increase in $\beta_4$ for the prolate minima of \nuc{74,76,78}{Kr}. The results obtained with the functional SkM*, which predicts prolate ground states with large quadrupole deformation $\beta_2 = 0.46 - 0.49$ for \nuc{74,76}{Kr} [see Fig.\,\ref{fig:sys}(a)], most closely resemble the observed experimental trend for the positive $\beta_4$ coupled-channels solution; also in absolute magnitude [see Fig.\,\ref{fig:sys}(b)]. In fact, Cl{\'e}ment {\it et al.} calculated a $\beta_2$ value of $\sim 0.5$ for the unperturbed prolate configuration in \nuc{74,76}{Kr} using two-state mixing with an unperturbed oblate configuration \cite{Cle07a}. This value is in excellent agreement with our DFT predictions for the prolate minimum [see Fig.\,\ref{fig:sys}\,(a)]. We note that UNEDF1, DD-PC1, and NL3* yield an oblate ground state for \nuc{74}{Kr}. In conflict with both coupled-channels solutions,  small $\beta_4$ values of $\sim -0.02$ are predicted at the oblate configurations. The predictions made with SkM* for the oblate minimum are shown in Fig.\,\ref{fig:sys}\,(b). They do not describe the experimentally observed trend emphasizing that large, positive $\beta_4$ values are associated with the prolate minima.

\begin{figure}[t]
\centering
\includegraphics[width=1\linewidth]{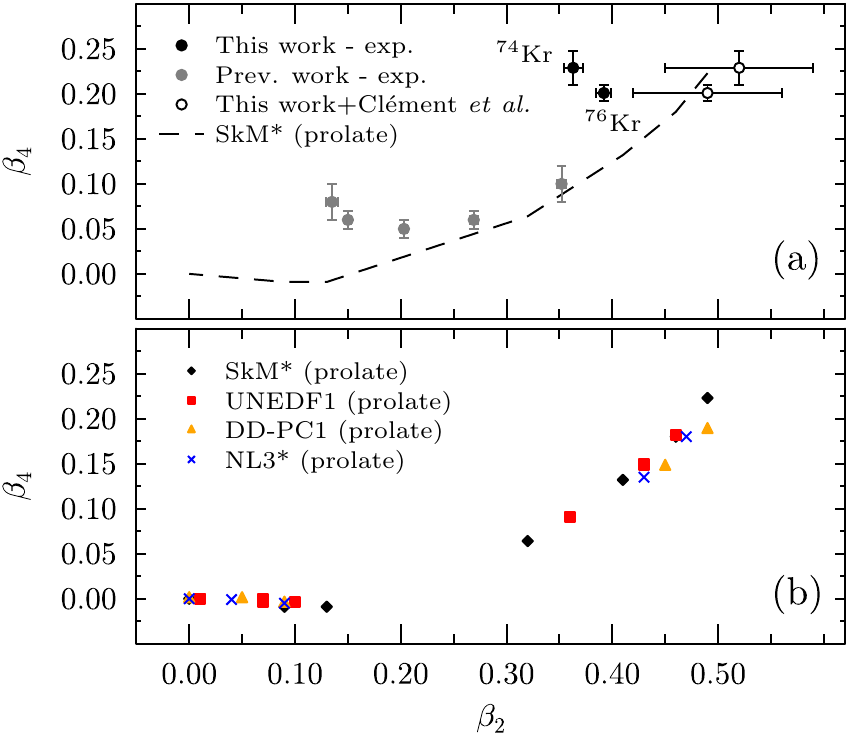}
\caption{\label{fig:b2b4}{Hexadecapole deformation parameter $\beta_4$  plotted against $\beta_2$: (a) 
experimental values;   (b) theoretical predictions. Earlier data for $\beta_4$ are from Ref.\,\cite{Sak79a}} (gray circles). The experimental $\beta_2$ values were taken from Ref.\,\cite{Pri14a, Pri16a}. The open circles in (a) correspond to the $\beta_2$ values calculated from the quadrupole moments $Q_{20}$ of the unperturbed prolate configurations reported in Ref.\,\cite{Cle07a}. The SkM* predictions are marked in  panel (a) with a dashed line.} 
\end{figure}

To show the connection between the large $\beta_2$ and  $\beta_4$ values, both the experimentally determined and theoretically predicted deformations are compiled in Fig.\,\ref{fig:b2b4}. It is seen that large values of $\beta_4$ correspond to  large values of $\beta_2$ both in experiment and theory. In fact, non-relativistic (SkM* and UNEDF1) and relativistic (DD-PC1 and NL3*) functionals conform to a remarkably similar trend [see Fig.\,\ref{fig:b2b4}\,(b)]. For nuclei with $\beta_2 \geq 0.3$, excellent agreement is observed between theory and experiment. The deviation of the experimental data for \nuc{74,76}{Kr} from this trend, when using the adopted $\beta_2$ values \cite{Pri14a, Pri16a}, further supports the shape-mixing hypothesis of Ref.\,\cite{Cle07a}. Considering that the $\beta_2$ values are smaller than 0.3 in the corresponding Ge and Se isotones \cite{Pri14a, Pri16a}, this connection also naturally explains why hexadecapole deformations  are significantly smaller \cite{Ogi86a, Sch87a}. For instance, the $N=40$ isotones of \nuc{76}{Kr}, \nuc{72}{Ge} [$\beta_2 = 0.240(2)$] and \nuc{74}{Se} [$\beta_2 = 0.290(8)$] have small $\beta_4$ values below 0.05. 

A simple interpretation of the positive hexadecapole moments can be obtained by using the geometric idea of the polar-gap model \cite{Ber68a} developed to explain  $\beta_4$ deformations of rare-earth nuclei \cite{Hen68a}. According to this model,
positive deformations $\beta_4$ are expected at the beginning of a shell and negative hexadecapole deformations are expected  at the end of a shell (see also Refs. \cite{Janecke1981,Moe16a,Naz81a, Ich86a, Ich87a, Cas00a}). 
The collective properties of $A=70-80$ nuclei are governed by deformed orbitals originating from the $0g_{9/2}$ unique-parity shell \cite{Naz85a}.
At large prolate quadrupole deformations $\beta_2 = 0.4 - 0.5$, the Nilsson levels [440]1/2 and [431]3/2 at the bottom of the $0g_{9/2}$ shell become occupied and this puts \nuc{74,76}{Kr} right at the beginning of the deformed shell with the low-$K$ Nilsson orbitals being at the Fermi surface \cite{Naz85a}. As also discussed in Ref.\,\cite{Cas00a}, these low-$K$ orbitals are primarily concentrated in the equatorial plane  of the quadrupole deformed core; hence, their population contributes to a positive hexadecapole moment. 

In summary, we have performed inelastic proton scattering experiments in inverse kinematics on the rare isotopes \nuc{74,76}{Kr} with GRETINA, the S800 spectrograph and NSCL/Ursinus LH$_2$ target. In this 
Letter, we report  $B(E4)$ strengths and $\beta_4$ deformations of these nuclei. Based on a
coupled-channels analysis,
two possible solutions for the $\beta_4$ hexadecapole deformation parameters of the $J^{\pi} = 4^+_1$ states in \nuc{74,76}{Kr} were determined, differing in sign because of the interference between one-step and two-step excitations. However, the two-step contributions are weak. Nuclear DFT calculations, employing both non-relativistic and relativistic energy density functionals, strongly  favor the solutions with large positive values of $\beta_4$. For this scenario, very good agreement between experiment and theory was obtained. This finding supports the 
results of Ref.\,\cite{Cle07a} suggesting predominantly prolate structures of the \nuc{74,76}{Kr} ground-, $J^{\pi} = 2^+_1$, and  $J^{\pi} = 4^+_1$ states. The large positive $\beta_4$ values reported in this Letter  are unambiguously linked to the prolate configurations. Given the remarkable  correlation between
$\beta_2$ and $\beta_4$ values predicted by our models for the Kr isotopes, it  suggests that $\beta_4$ deformations could indeed be a sensitive indicator of prolate shapes. Since the prolate-oblate ground-state phase transition presumably happens at $A=72$ (or $N=36$), it is a valid question what one might observe for the hexadecapole moments of \nuc{70,72}{Kr} and other nuclei in the neutron-deficient Ge-Sr region. Future experiments at rare isotope beam facilities will help answering this question.

\section*{Acknowledgements}
This work was supported by the National Science Foundation (NSF) under Grant No. PHY-2012522 (WoU-MMA: Studies of Nuclear Structure and Nuclear Astrophysics), Grant No. PHY-1565546 (NSCL), Grant No. PHY-2209429 (Windows on the Universe: Nuclear Astrophysics at FRIB),  by the Department of Energy, Office of Science, Office of Nuclear Physics, Grant Nos. DE-SC0020451 and DE-SC0013365 (MSU), and by the Department of Energy, NNSA, Grant No. DOE-DE-NA0004074 (the Stewardship Science Academic Alliances program). GRETINA was funded by the Department of Energy, Office of Science. The operation of the array at NSCL was supported by the DOE under Grant No. DE-SC0019034. M.S. acknowledges support through the FRIB Visiting Scholar Program for Experimental Science 2020.






\bibliographystyle{apsrev4-1}
\bibliography{Kr_hexa.bib}

\end{document}